\documentclass[conference]{IEEEtran}
\IEEEoverridecommandlockouts
\usepackage{cite}
\usepackage{amsmath,amssymb,amsfonts}
\usepackage{algorithmic}
\usepackage{graphicx}
\usepackage{textcomp}
\usepackage{xcolor}
\usepackage{tikz}
\usepackage{pifont}
\usepackage{comment}
\usepackage{array}
\usepackage{todonotes}
\usepackage{hyperref}

\newcommand*\circled[1]{\tikz[baseline=(char.base)]{
            \node[shape=circle,draw,inner sep=1.3pt] (char) {#1};}}

\def\BibTeX{{\rm B\kern-.05em{\sc i\kern-.025em b}\kern-.08em
    T\kern-.1667em\lower.7ex\hbox{E}\kern-.125em}}

\usepackage{moresize}
\usepackage{listings}

\definecolor{darkgreen}{rgb}{0.0,0.4,0.0} 
\definecolor{darkred}{rgb}{0.6,0.1,0.1}
\definecolor{lightgray}{gray}{.98}
\definecolor{medgray}{gray}{.70}
\definecolor{darkgray}{gray}{.40}
\definecolor{lightviolet}{rgb}{0.7,0,0.7} 
\definecolor{lightlightviolet}{rgb}{1.0,0.7,1.0} 
\definecolor{darkviolet}{rgb}{0.5,0.1,0.5}
\definecolor{darkredviolet}{rgb}{0.6,0.1,0.4}
\definecolor{limegreen}{rgb}{0.2,0.7,0.2}
\definecolor{navyblue}{RGB}{0,0,128}
\definecolor{aquamarine}{RGB}{102,205,170}

\definecolor{strictRED}{RGB}{184,0,0}
\definecolor{specificationTURQUOISE}{RGB}{0,128,153}
\definecolor{assumptionGREEN}{RGB}{0,128,0}
\definecolor{interruptBLUE}{RGB}{0,0,128}
\definecolor{committedORCHID}{RGB}{54,22,89}
\definecolor{urgentORCHID}{RGB}{74,28,109}
\definecolor{requestedORCHID}{RGB}{104,34,139}
\definecolor{eventuallyORCHID}{RGB}{154,50,205}

\lstdefinelanguage{SMLX}
{
	basicstyle=\ssmall\ttfamily, %
	frame=single, 
	framextopmargin=0pt,
	framexbottommargin=0pt,
	framexleftmargin=0pt,
	xleftmargin=16pt,
	xrightmargin=3pt,
	morekeywords=[1]{system, domain, scenario, bind, to, 
		message, non, spontaneous, events, specification, 
		alternative, if, collaboration, role, with, dynamic, 
		bindings, or, and, null, define, as, 
		constraints, import, static, parameter, ranges, var, EInt, 
		controllable},
	morekeywords=[2]{strict},
	morekeywords=[3]{forbidden, violation},
	morekeywords=[4]{interrupt},
	morekeywords=[5]{guarantee},
	morekeywords=[6]{assumption}, 
	morekeywords=[7]{committed}, 
	morekeywords=[8]{urgent},
	morekeywords=[9]{requested},
	morekeywords=[10]{eventually},
	keywordstyle=[1]\color{darkviolet}\textbf,
	keywordstyle=[2]\color{strictRED}\textit,
	keywordstyle=[3]\color{strictRED}\textit,
	keywordstyle=[4]\color{interruptBLUE}\textit,
	keywordstyle=[5]\color{specificationTURQUOISE}\textbf,
	keywordstyle=[6]\color{assumptionGREEN}\textbf,
	keywordstyle=[7]\color{committedORCHID}\textit,
	keywordstyle=[8]\color{urgentORCHID}\textit,
	keywordstyle=[9]\color{requestedORCHID}\textit,
	keywordstyle=[10]\color{eventuallyORCHID}\textit,
	sensitive=false,
	morecomment=[l][\color{darkgreen}\textit]{//},
	morecomment=[s][\color{darkgreen}\textit]{/*}{*/}, 
	morestring=[b][\color{blue}]",
	tabsize=1,
	moredelim = [s][\color{specificationTURQUOISE}\textbf]{guarantee}{scenario},
	moredelim = [s][\color{assumptionGREEN}\textbf]{assumption}{scenario},
	backgroundcolor=\color{lightgray}
}

\lstdefinestyle{SMLXStyle} {language=SMLX}

\lstdefinelanguage{SMLConfig}
{
	basicstyle=\ssmall\ttfamily,
	frame=single, 
	framextopmargin=0pt,
	framexbottommargin=0pt,
	framexleftmargin=0pt,
	xleftmargin=16pt,
	xrightmargin=3pt,
	morekeywords=[1]{symbolic, import, configure, specification, use, 
		instancemodel, symbolic, parameters, attributes, symbolic, state, matching, 
		off, under, approximation, on, rolebindings, collaboration, object, plays, 
		role, role1},
	keywordstyle=[1]\color{darkviolet}\textbf,
	sensitive=false,
	morecomment=[l][\color{darkgreen}\textit]{//},
	morecomment=[s][\color{darkgreen}\textit]{/*}{*/}, 
	morestring=[b][\color{navyblue}\textit]",
	stringstyle=\color{navyblue},
	tabsize=1,
	backgroundcolor=\color{lightgray}
}

\lstdefinestyle{SMLConfigStyle} {language=SMLConfig}
\lstdefinelanguage{Java}
{
	basicstyle=\ssmall\ttfamily,
	frame=single, 
	framextopmargin=0pt,
	framexbottommargin=0pt,
	framexleftmargin=0pt,
	xleftmargin=16pt,
	xrightmargin=3pt,
	morekeywords=[1]{public, private, class, extends, protected, void,
		new, throws, null, if, else},
	morekeywords=[2]{STRICT},
	morekeywords=[3]{@Override},
	morekeywords=[4]{car, oc, cp}, 
	keywordstyle=[1]\color{darkviolet}\textbf,
	keywordstyle=[2]\color{javablue}\textbf,
	keywordstyle=[3]\color{darkgray},
	keywordstyle=[4]\color{navyblue},
	sensitive=false,
	morecomment=[l][\color{javagreen}\textit]{//},
	morecomment=[s][\color{javagreen}\textit]{/*}{*/}, 
	morestring=[b][\color{javablue}\textit]",
	stringstyle=\color{navyblue},
	tabsize=1,
	backgroundcolor=\color{lightgray}
}

\definecolor{javared}{rgb}{0.6,0,0} 
\definecolor{javablue}{rgb}{0,0,0.9} 
\definecolor{javagreen}{rgb}{0.25,0.5,0.35} 
\definecolor{javapurple}{rgb}{0.5,0,0.35} 
\definecolor{javadocblue}{rgb}{0.25,0.35,0.75} 

\lstdefinestyle{JavaStyle} {language=Java}

\lstdefinelanguage{Kotlin}{
	basicstyle=\ssmall\ttfamily,
	frame=single, 
	framextopmargin=0pt,
	framexbottommargin=0pt,
	framexleftmargin=0pt,
	xleftmargin=16pt,
	xrightmargin=3pt,
	comment=[l]{//},
	commentstyle={\color{darkgray}\ttfamily},
	emph={delegate, filter, first, firstOrNull, forEach, lazy, map, mapNotNull, println, return@, event, sends, request, requestParamValuesMightVary},
	emphstyle={\color{darkviolet}},
	identifierstyle=\color{black},
	numberstyle=\color{darkgreen},
	keywords=[1]{ abstract, actual, as, as?, break, by, companion, continue, data, do, dynamic, else, enum, expect, false, final, for, get, if, import, in, interface, internal, is, null, object, override, package, private, public, return, set, super, suspend, this, throw, true, try, typealias, val, var, vararg, when, where, while},
	keywordstyle=[1]{\color{javablue}\bfseries},
	keywords=[2]{@Deprecated, @JvmField, @JvmName, @JvmOverloads, @JvmStatic, @JvmSynthetic, @Test, Array, Byte, Double, Float, Int, Integer, Iterable, Long, Short, String, scenario, class, fun},
	keywordstyle=[2]{\color{javablue}},	
	keywords=[3]{interruptingEvents, forbiddenEvents, it}, 
	keywordstyle=[3]{\color{darkviolet}\bfseries},
	keywords=[4]{scenario, cycleScenario, runTest, Given, When, Then, And, But}, %
	keywordstyle=[4]{\textit},
	keywords=[5]{coolantTemp, deratingFactor}, 
	keywordstyle=[5]{\color{darkviolet}\bfseries\underbar},
	keywords=[6]{currentTemp}, 
	keywordstyle=[6]{\underbar},
	morecomment=[s]{/*}{*/},
	morecomment=[s][\color{black}]{`}{`},
	morestring=[b]",
	morestring=[s]{"""*}{*"""},
	sensitive=true,
	stringstyle={\color{javagreen}\ttfamily},
}

\lstdefinestyle{KotlinStyle} {language=Kotlin}

\newcommand{\lstinlineKotlin}[1]{\lstinline[language=Kotlin,basicstyle=\small\ttfamily]{#1}}

\lstdefinelanguage{Gherkin}{
	basicstyle=\ssmall\ttfamily,
	frame=single, 
	framextopmargin=0pt,
	framexbottommargin=0pt,
	framexleftmargin=0pt,
	xleftmargin=16pt,
	xrightmargin=3pt,
	comment=[l]{//},
	commentstyle={\color{darkgray}\ttfamily},
	emph={@software, @charging },
	emphstyle={\color{limegreen}},
	identifierstyle=\color{black},
	keywords={Given, When, Then, And},
	keywordstyle={\color{violet}\ttfamily},
	morecomment=[s]{/*}{*/},
	morestring=[b]",
	morestring=[s]{"""*}{*"""},
	ndkeywords={Scenario, Example, Feature},
	ndkeywordstyle={\color{darkviolet}\bfseries},
	sensitive=true,
	stringstyle={\color{javagreen}\ttfamily},
}

\lstdefinestyle{GherkinStyle} {language=Gherkin}

\makeatletter
\def\ps@IEEEtitlepagestyle{%
  \def\@oddfoot{\mycopyrightnotice}%
  \def\@evenfoot{}%
}
\def\mycopyrightnotice{%
  {\begin{minipage}{\textwidth}
  \footnotesize \copyright 2021 IEEE. Personal use of this material is permitted. Permission from IEEE must be obtained for all other uses, in any current or future media, including reprinting\slash republishing this material for advertising or promotional purposes, creating new collective works, for resale or redistribution to servers or lists, or reuse of any copyrighted component of this work in other works.
  \end{minipage}
  }
  \gdef\mycopyrightnotice{}
}
\lstdefinestyle{mystyle}{
	basicstyle=%
	\ttfamily
	\lst@ifdisplaystyle\footnotesize\fi
}
\makeatother

\begin{document}

\title{Integrated and Iterative Requirements Analysis\\ and Test Specification: A Case Study at Kostal}

\author{
    \IEEEauthorblockN{
    Carsten Wiecher\IEEEauthorrefmark{1}, 
    Jannik Fischbach\IEEEauthorrefmark{2},
    Joel Greenyer\IEEEauthorrefmark{3}, 
    Andreas Vogelsang\IEEEauthorrefmark{4},
    Carsten Wolff\IEEEauthorrefmark{5},
    Roman Dumitrescu\IEEEauthorrefmark{6}}
    \IEEEauthorblockA{\IEEEauthorrefmark{1}\textit{KOSTAL Automobil Elektrik GmbH \& Co. KG, 44227 Dortmund, Germany}}
    \IEEEauthorblockA{\IEEEauthorrefmark{2}\textit{Qualicen GmbH, 85748 Munich, Germany
    }}
    \IEEEauthorblockA{\IEEEauthorrefmark{3}\textit{FHDW Hannover, 30173 Hannover, Germany}
    }
    \IEEEauthorblockA{\IEEEauthorrefmark{4}\textit{University of Cologne, 50923 Cologne, Germany
    }}
    \IEEEauthorblockA{\IEEEauthorrefmark{5}\textit{Dortmund University of Applied Sciences and Arts, 44139 Dortmund, Germany
    }}
    \IEEEauthorblockA{\IEEEauthorrefmark{6}\textit{Fraunhofer IEM, 33102 Paderborn, Germany
    }}
    c.wiecher@kostal.com, jannik.fischbach@qualicen.de, joel.greenyer@fhdw.de, vogelsang@cs.uni-koeln.de,\\ carsten.wolff@fh-dortmund.de, roman.dumitrescu@iem.fraunhofer.de
}

\maketitle

\begin{abstract}
Currently, practitioners follow a top-down approach in automotive development projects. However, recent studies have shown that this top-down approach is not suitable for the implementation and testing of modern automotive systems. Specifically, practitioners increasingly fail to specify requirements and tests for systems with complex component interactions (e.g., e-mobility systems). In this paper, we address this research gap and propose an integrated and iterative scenario-based technique for the specification of requirements and test scenarios. Our idea is to combine both a top-down and a bottom-up integration strategy. For the top-down approach, we use a behavior-driven development (BDD) technique to drive the modeling of high-level system interactions from the user's perspective. For the bottom-up approach, we discovered that natural language processing (NLP) techniques are suited to make textual specifications of existing components accessible to our technique. To integrate both directions, we support the joint execution and automated analysis of system-level interactions and component-level behavior. We demonstrate the feasibility of our approach by conducting a case study at Kostal (Tier1 supplier). The case study corroborates, among other things, that our approach supports practitioners in improving requirements and test specifications for integrated system behavior.
\end{abstract}


\section{Introduction}
\textbf{Context} 
A key electronic control unit (ECU) in battery electric vehicles (BEVs) is the on-board charger (OBC), which acts as a communication gateway for the charging infrastructure and converts energy to charge the BEVs' battery~\cite{Schnitzler2020}.
In the OBC development context, we especially focus on challenges related to requirements-based testing, as this is the state of practice in the automotive domain~\cite{ISO26262_2018,Liebel2018,Liebel2019}. 
Currently, practitioners follow a top-down approach when developing and testing automotive systems~\cite{Penzenstadler2010,Bohm2014,Braun2014a}. Specifically, they first decompose high-level requirements into more granular requirements that can be implemented. Subsequently, the requirements specifications on the system, sub-system, and component level are used to create test specifications, which are utilized to verify if the requirements were implemented correctly and to validate if the system conforms with the stakeholder requirements, i.e., using component- and integration tests~\cite{Kasauli2021,Juhnke2020}.      

\textbf{Problem} However, this top-down and linear requirements decomposition and test specification is difficult to follow in an e-mobility context~\cite{Kirpes2019}, where the functionality is realized by integrating systems that have different life cycles and are developed in different organizations 
(e.g., charging stations, electric vehicle supply equipment, smartphone apps) \cite{Kasauli2021}. 

\textbf{Goal} In this paper, we address this research gap and focus on the question of \textit{how to best support the specification of requirements and tests for systems with complex component interactions, and with a strong focus on component reuse}.

\textbf{Contributions} Following the design science research approach~\cite{Dresch2015}, we developed an iterative and integrated specification method (see Fig.~\ref{fig:method}). Our idea is to use executable and scenario-based requirements models, combined with intuitive requirements simulation techniques~\cite{Harel2003}, to support practitioners in the integrative specification of requirements and tests. Specifically, we combine the existing top-down strategy with a bottom-up approach and build on our previously devised test-driven scenario specification method (TDSS)~\cite{Wiecher2019}, where tests are used to drive the iterative formalization of functional requirements, in order to automatically analyze the system behavior. As a bottom-up integration strategy, we use natural language processing (NLP) techniques~\cite{Fischbach2020} to generate tests from component requirements. As a top-down strategy, we use the behavior-driven development (BDD) paradigm~\cite{Wiecher2020} to generate tests from stakeholder expectations represented via usage scenarios. 
We integrate both directions to create suitable integration tests and to validate if the stakeholder expectations can be met when the detailed behavior of existing specifications is considered. In this way, we support the estimation of the extent to which existing requirements can be reused when stakeholder expectations change. 


\textbf{Practical Impact} We demonstrate the feasibility of our approach in a case study at a Tier1 supplier company. We found that:
\begin{itemize}
    \item We were able to automatically generate test cases based on existing textual requirements specifications, which we used as input for our TDSS method when following the bottom-up integration strategy.  
    The majority (71\%) of manually designed test cases could be generated automatically by our approach. Interestingly, we were able to generate 22 test cases that were not considered in the manually created set. For the 59 test cases that could not be generated automatically, we determined missing information or grammatical errors in the requirements specification. 
    \item Following the top-down integration strategy, we were able to successfully apply the BDD approach to model component interactions. We found that acceptance criteria formulated as usage scenarios are suitable to drive the scenario-based specification of end-to-end event chains of an integrated system. Based on the scenario specification, the requirements simulation and visualization in the form of generated sequence diagrams support both requirements and test specifications in development projects in our case company.  
    \item By integrating both strategies, we were able to create suitable integration test scenarios, which can be used to trigger the component interaction based on the top-down approach, while also considering detailed requirements coming from the bottom-up approach. In this way, the test scenario covers already existing component requirements and creates a link to (changing) stakeholder needs. 
\end{itemize}

\section{Background}
\subsection{On-Board Charger for Battery Electric Vehicles}
\label{sect:obc}
Within BEVs, the OBC is used to convert AC voltage from the grid into DC voltage of the vehicle's battery. It´s expected that OBCs are going to be a commodity component of electric vehicles within the next years~\cite{Schnitzler2020}. Consequently, to satisfy the needs of a global market, we require efficient charging at different power grids in different regions (e.g. three-phase 11kW grids in Europe, or single phase charging at 7.2kW in China). In addition, different charging standards (e.g. CHAdeMO \cite{CHAdeMOassociation2012}, Chinese GB/T DC standard~\cite{GB/T20234}) must be supported when the OBC is acting as a communication gateway. This leads to high system complexity, even for a single ECU~\cite{Schnitzler2020}. In this development context, it is important to ensure the functional safety of high voltage components. Hence, the specification of requirements and related specifications of tests are key artifacts of system engineering activities.    

\subsection{Automotive Requirements Specification}
\label{sect:requirementsSpecification}
The state of practice for the development of safe and secure automotive systems is based on top-down requirements decomposition approaches~\cite{Penzenstadler2010,Bohm2014,Braun2014a} and is guided by standards like ASPICE~\cite{AutomotiveSIG2015}, ISO 26262~\cite{ISO26262_2018}, and ISO/SAE 21434~\cite{Schmittner2018}, to ensure a continuous traceability~\cite{Holtmann2020} from initial system requirements to validated development artifacts. In this context, several model-based requirements engineering approaches exist that aim to make the specification and decomposition of requirements more efficient (e.g. \cite{Braun2014a, Bohm2014, Holtmann2016}). However, these approaches are rarely applied in industry, since they are not suited for the development of next-generation automotive systems~\cite{Liebel2019,Ncube2018,Hoehne2018}.
Recent studies reveal that traditional top-down requirements engineering approaches fail to cover the increasing connectivity of modern vehicles, which can be seen as a constituent system in a system of systems~\cite{Lane2013WhatIA,Nielsen2015} (e.g. smart charging in an e-mobility systems context~\cite{Kirpes2019}). 
We addressed this research gap in previous work and proposed a scenario-based requirements modeling approach to support iterating, changing, synchronizing, and communicating requirements across different abstraction and hierarchy levels as well as scopes of responsibility~\cite{Wiecher2021, Wiecher2021b}. In this paper, we use these concepts to model component interactions and functional component requirements to support the specification of requirements in natural language that provide a "common language" for engaging non-technical stakeholders in the system development process~\cite{Liebel2018}, which in the automotive industry is typically done with requirements lists, supported with tools like \emph{IBM Doors}~\cite{IBM2020}.    

\subsection{Scenario-based modeling of functional requirements}
\label{sect:smlk}
We use the scenario modeling language for Kotlin (SMLK) to formalize functional requirements. SMLK is a Kotlin-based framework that leverages the concepts of Live Sequence Charts (LSCs)~\cite{Damm2001}, the Behavioral Programming (BP) paradigm~\cite{Harel2012}, and the Scenario Modeling Language~\cite{Greenyer2017}. An SMLK program consists of \emph{scenarios}, which are threads that capture individual behavioral aspects that are loosely coupled via shared events. When executing scenarios as a \emph{scenario program}, the scenarios are interwoven to achieve a consistent system behavior that satisfies the requirements of all scenarios.   

For a feasible formalization of requirements, we make use of the TDSS process~\cite{Wiecher2019}. Motivated by test-driven development, we use tests to trigger the execution of our scenario specification.
Thereby, we follow the iterative process as outlined in Fig.~\,\ref{fig:tdss-process}:
\begin{figure}[h]
    \centering
    \includegraphics[width=1.0\linewidth]{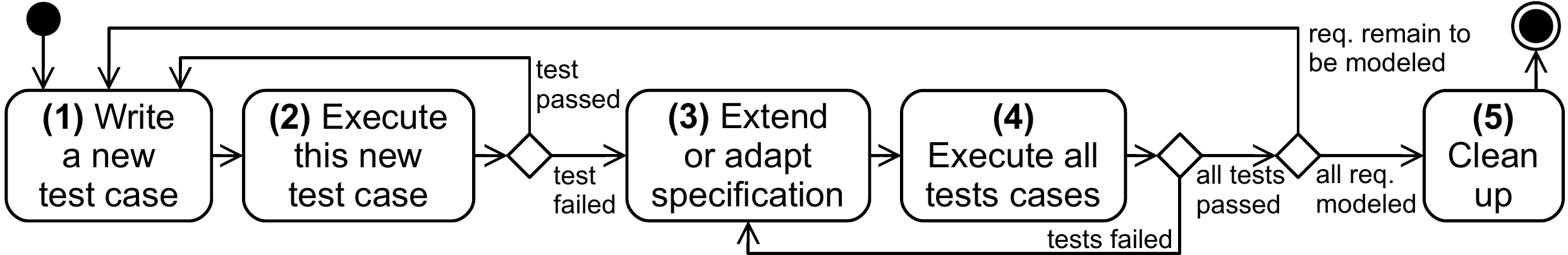}
    \caption{Test-Driven Scenario Specification (TDSS)~\cite{Wiecher2019}}
    \label{fig:tdss-process}
\end{figure}
We first take a requirement and write a test case for it. When executing this test case it is expected to fail. We then add scenarios to our scenario specification to satisfy all tests. In this way, we iteratively create executable requirements and test specifications and reduce the risk of ambiguities and inconsistencies in requirements~\cite{Wiecher2019}.

\begin{figure}[h]
    \centering
    \includegraphics[width=1.0\linewidth]{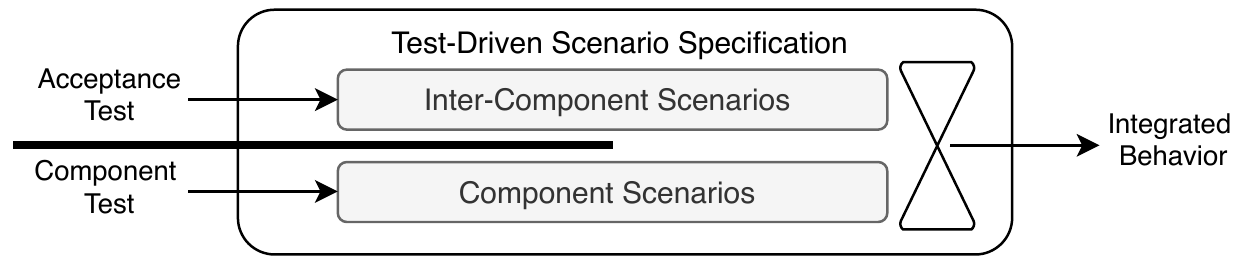}
    \caption{Test-driven scenario specification on different hierarchy levels.}
    \label{fig:intervscomponent}
\end{figure}

To consider the different integration strategies in our approach, we use TDSS on two different hierarchy levels, which we support by using two scenario types as outlined in Fig.~\ref{fig:intervscomponent}:
We use the term \emph{inter-component scenario} to model requirements of how the components of an ECU software system shall interact with external systems and each other in order to realize the overall system functionality. 
We use \emph{component scenarios} to model the requirements of how every single component within that system shall react to events from other components. 

As an example, Listing~\ref{list:interComponentScenario} specifies an inter-component scenario that is triggered when the hardware control component of the OBC receives a control pilot signal from the charging socket of the vehicle. This is modeled by an interaction event of the \lstinlineKotlin{chargingSocket} object sending the message \lstinlineKotlin{cpSignalHW} to the \lstinlineKotlin{hardwareControl} object (line~1). In the scenario body, we model the interaction between the different components which are necessary to process the control pilot signal (line 3 and 4).
The execution of this scenario can be triggered by a test as shown in Listing \ref{list:interComponentScenarioTest}, where we execute the trigger event in (line 3), and subsequently expect to receive the events that are requested in the inter-component scenario body (line 3 and~4 in Listing \ref{list:interComponentScenario}). 
\begin{lstlisting}[caption=Test case that triggers the inter-component scenario execution,
	label=list:interComponentScenarioTest,
	style=KotlinStyle
	]
@Test
fun testCpSignalProcessing (){
trigger(chargingSocket sends hardwareControl.cpSignalHW(0.0))
receive(hardwareControl sends controlPilot receives ControlPilot::cpSignalSW)
receive(controlPilot sends application receives Application::cpSignalInformation)}
\end{lstlisting}

To model the behavior, we use different programming idioms~\cite{Harel2012}. In addition to the existing idioms, we use the scenario method \lstinlineKotlin{requestParamValuesMightVary} that allows us to request events with supplied default parameter values, that can be overridden later (i.e., if other scenarios request the same events with different parameter values, the scenario will also accept, and progress on, such events~\cite{Wiecher2021}). In this way, we abstract from component details, but we already can execute the inter-component specification of the overall system as a scenario program.
This is useful when modeling on a level of abstraction, where possibly not all parameter values are already known, as they might be based on concrete sensor readings or run-time parameters that we do not want to specify in detail on this level.
\begin{lstlisting}[caption=Inter-component scenario,
	label=list:interComponentScenario,
	style=KotlinStyle
	]
scenario(chargingSocket sends hardwareControl receives HardwareControl::cpSignalHW){
    val cpSignalValue = it.parameters[0] as Double
    requestParamValuesMightVary(hardwareControl sends controlPilot.cpSignalSW("*", cpSignalValue))
    requestParamValuesMightVary(controlPilot sends application.cpSignalInformation("*"))}
\end{lstlisting}

Listing~\ref{list:intraComponentScenario} shows a component scenario to model the control pilot behavior. Component specifications can also be created using the TDSS approach. Moreover, the component specification scenarios can also be executed together with the inter-component specification scenarios in order to validate whether the specifications on the two hierarchy levels are consistent (cf.~\cite{Wiecher2021b}).
In a joint execution of component- and inter-component scenarios, for example, the interaction event  \lstinlineKotlin{hardwareControl sends controlPilot.cpSignalSW("*", cpSignalValue)}, requested in the inter-component scenario (line 3 in Listing~\ref{list:interComponentScenario}), may be executed, and will trigger the component scenario shown in Listing \ref{list:intraComponentScenario}, where the component behavior is specified in more detail. Specifically, we modeled that an output value shall be set if the control pilot signal status is \lstinlineKotlin{RTE_E_OK} and the signal value is \lstinlineKotlin{3.0}. Eventually (see line 6), we request an interaction event with concrete parameter values. 
\begin{lstlisting}[caption=Component scenario for the control pilot component,
	label=list:intraComponentScenario,
	style=KotlinStyle
	]
scenario(hardwareControl sends controlPilot receives ControlPilot::cpSignalSW) {
    val status = it.parameters[0] as String
    val value = it.parameters[1] as Double
    if ((status == "RTE_E_OK") && (value == 3.0)) {
        request(controlPilot.setOutputValue("READY_WITH_VENT", value))
        request(controlPilot sends application.cpSignalInformation("READY_WITH_VENT"))
    ...        
}
\end{lstlisting}

\subsection{Automotive Test Case Specification}
\label{sect:testCaseSpecification}
\begin{figure*}
    \centering
    \includegraphics[width=1.0\linewidth]{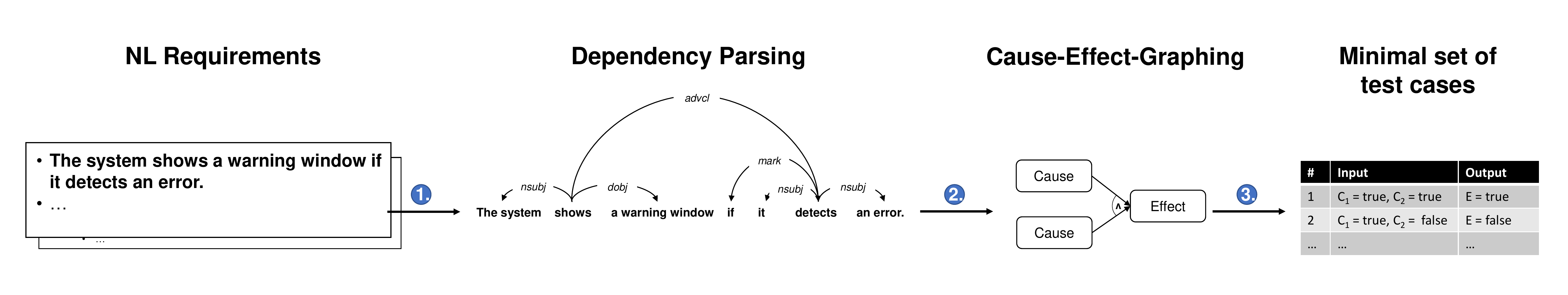}
    \vspace{-.7cm}
    \caption{Overview of the \emph{Specmate} pipeline~\cite{Fischbach2020}.}
    \label{fig:specmate_pipeline}
\end{figure*}
Verification and validation are crucial activities in the development of automotive systems~\cite{Lindemann2016}. They are used to align the state of development with stakeholder objectives and to evaluate if the development artifacts comply with safety and security goals as prescribed by automotive standards~\cite{ISO26262_2018,Schmittner2018} and maturity models~\cite{AutomotiveSIG2015}. These standards require mandatory documentation of work products~\cite{Juhnke2020}. One artifact in this context is the \emph{test case specification} that includes a set of test cases that are derived from a \emph{test basis} (e.g. requirements specification) for a defined \emph{test object} (e.g implemented software- or hardware/software system) \cite{Spillner2011,AutomotiveSIG2015,Juhnke2020}.  
In the automotive domain, the creation of the test case specification can be seen as a sub-process in the overall system development process and can be separated into the phases 1) test basis analysis 2) test case design \& distribution 3) test case documentation, and 4) test case specification review \& approval~\cite{Spillner2011, Coordinating2013}. To support the test designer to execute these process steps and to create high-quality test case specifications, we propose to use executable and scenario-based requirements models to iteratively specify test cases (as already started in previous work~\cite{Wiecher2019, Wiecher2020, Wiecher2020a}). In this paper, we extend our previous work and use \emph{Specmate}~\cite{Fischbach2020} to automatically derive test cases from natural language requirements, which in turn are used to iteratively create scenario-based test specifications. The resulting artifact of our approach is a \emph{system integration test specification}.

\subsection{Automated Creation of Test Cases with  \emph{Specmate}}
\label{sect:specmate}
\emph{Specmate} is an approach to automatically create test cases for natural language (NL) requirements~\cite{Fischbach2020}. \emph{Specmate} follows the idea of Model-Based-Testing and introduces an intermediate layer between the NL requirements and the final test cases. Consequently, each requirement is first transferred into a test model from which we then derive the right amount of test cases. Since functional requirements specify the expected system behavior often in the form of cause-effect relationships (e.g., In the case of \textit{cause}, the system shall \textit{effect})~\cite{fischbachREFSQ}, we use Cause-Effect-Graphs (CEG) as test models. As illustrated in Fig.~\ref{fig:specmate_pipeline}, the \emph{Specmate} pipeline consists of the following three steps:

Firstly, the syntax and semantics of a causal requirement are analyzed by applying a Dependency Parser. We use the Malt Parser (Version 1.9.2) to create a dependency tree representing the overall structure of the requirement. A dependency tree consists of a set of subtrees, which represent different parts of the parsed sentence. We aim to identify those subtrees that comprise the individual causes and effects. For this purpose, we traverse the tree and identify individual subtrees by pattern matching. Specifically, \emph{Specmate} checks which fragments in the sentence conform to one of our 38 pre-defined dependency patterns and extracts them accordingly. Due to the brevity of space, we refer to our previous paper~\cite{Fischbach2020} for a more detailed explanation of the pattern matching. In the second step, the extracted causes and effects are mapped to a Cause-Effect-Graph. To derive test cases from the CEG, \emph{Specmate} applies the Basic Path Sensitization Technique (BPST)~\cite{Nursimulu95}. The graph is traversed back from the effects to the causes and test cases are created according to specific decision rules, which can be found in~\cite{source6}. These rules achieve the maximum probability of finding failures while avoiding the complexity of generating 2\textsuperscript{n} test cases, where n is the number of causes. Thus, \textit{Cause-Effect-Graphing} enables balancing between sufficient test coverage and the lowest possible number of test cases.

\section{Integrated and Iterative Requirements Analysis and Test Specification}
\label{sect:approach}
\begin{figure*}[h]
    \centering
    \includegraphics[width=1.0\linewidth]{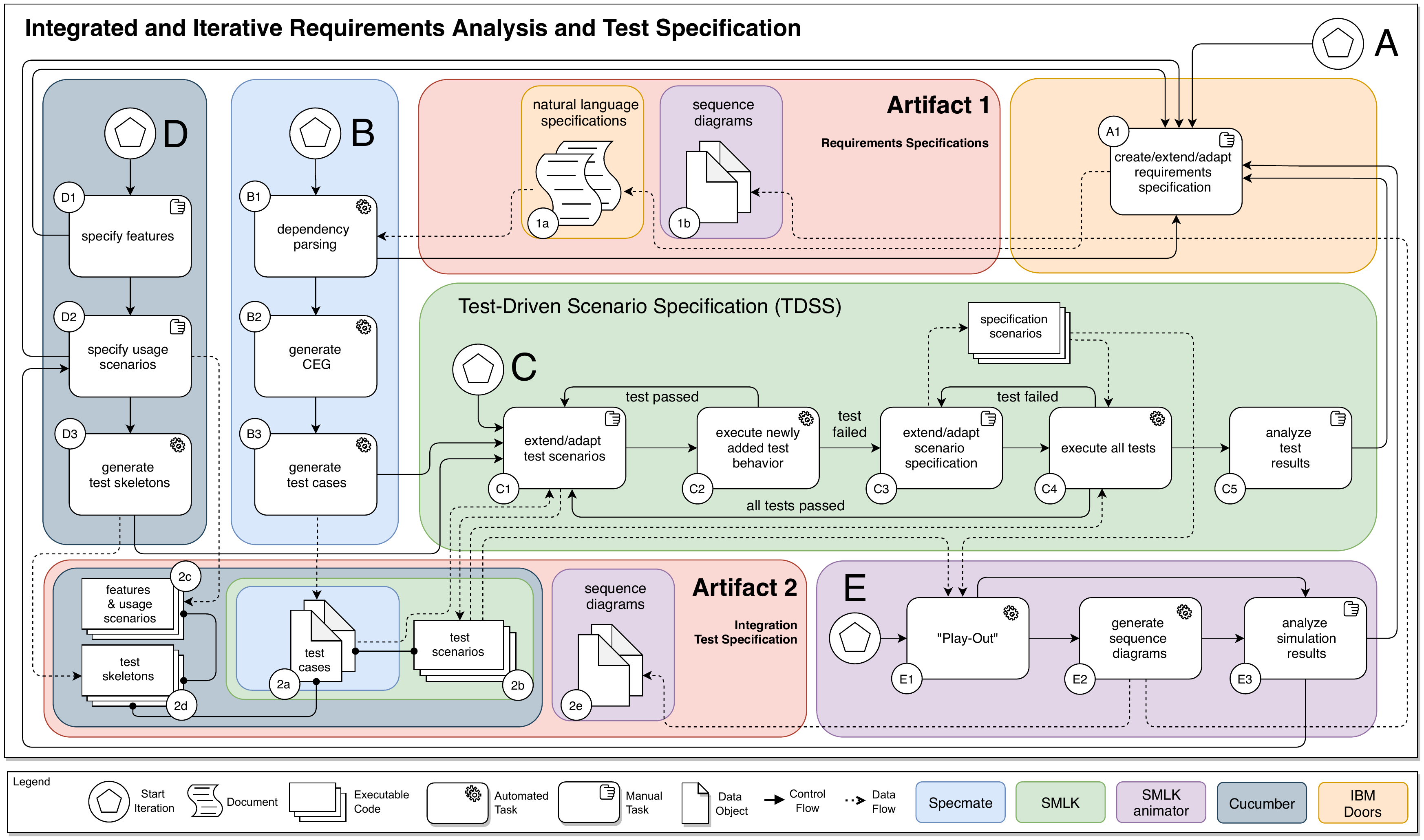}
    \caption{Closed loop process steps for the integrated and iterative requirements analysis and test specification.}
    \label{fig:method}
\end{figure*}

In this paper, we address the question of \textit{how to best support the specification of requirements and tests for systems with complex component interactions, and with a strong focus on component reuse}. This section outlines our proposed solution: an integrated and iterative method for the specification and analysis of requirements and tests. Fig.~\ref{fig:method} provides an overview of our approach, which consists of five sub-processes (A-E). Since the available data and development scope changes constantly during modern development projects, we designed a hybrid bottom-up and top-down specification approach with several entry points. Hence, the five sub-processes can be started separately. However, the single process outcomes like \emph{features}, \emph{usage scenarios}, \emph{test cases}, \emph{system scenarios} etc. are highly related to each other with the goal to support the creation of two main artifacts: 1) a \emph{requirements specification}, and 2) an \emph{integration test specification}. In the following, we describe how these artifacts are created by iterating through the sub-processes and describing their dependencies.  

\circled{A} 
One entry point is the specification of requirements in natural language, which in our case is done using \emph{IBM Doors}~\cite{IBM2020} to allow the linking of requirements and ensure the traceability of requirements within the overall development process. If a specification already exists, the feedback from the dependency parsing (B1), the feature specification (D1, D2), and the requirements validation \& verification (C5, E3) is used to extend or adapt the existing requirements specification. 

\circled{B}
If requirements are already available, we use \emph{Specmate}~\cite{Fischbach2020} to automatically detect causal dependencies in these requirements (B1). Based on the identified causal relations we create cause-effect graphs (CEGs) (B2) that allow us to automatically generate the optimal amount of test cases for each requirement (B3), as introduced in Sect.~\ref{sect:specmate}. We recommend starting the specification process with sub-process B when following a bottom up-integration strategy, where we already have a high amount of requirements that describe the behavior of single components. 

\circled{C}
The generated test cases are input for the TDSS process: In C1, we manually add \emph{SMLK} events to the generated tests to enable automatic execution of the newly added tests in C2.
According to the test-driven development paradigm, these tests are expected to fail until we model the respective requirement as an executable scenario (see C3). Subsequently, we continue with several iterations until all tests pass in C4. The motivation of this iterative process is to provide a feasible approach for a step-wise formalization of requirements. 

\circled{D}
When running through the sub-processes A-C by using real-world specifications, we have to process a high amount of requirements (see exemplary Sect.~\ref{sect:casestudy} for a single component). Consequently, when iteratively modeling all scenarios considering each requirement, this can lead to a loss of reference to the actual problem (cf. to the sampling concern in scenario-based requirements engineering \cite{Sutcliffe2003}). 
To address this, we extended the test case generation in B and the test-driven scenario specification in C with the BDD approach in D, to create a clear reference to the stakeholder need and support a top-down integration strategy. Within the sub-process D, we use the BDD tool Cucumber~\cite{cucumber} in order to create \emph{feature specifications} (D1) that include \emph{usage scenarios} (D2)~\cite{SmartBearSoftware} representing stakeholder expectations. We use these stakeholder expectations to generate test skeletons (D3), which we then apply to structure the single test cases that we generated in B3. By integrating Cucumber with SMLK, we are able to execute single features that in turn trigger the execution of the structured test cases. In this way, the BDD approach supports the test designer to combine single test cases to integration test sequences that have a link to the stakeholder need. 

We integrate the BDD tool Cucumber with SMLK due to the following reasons: Firstly, it allows us to support test designers to create valuable test scenarios with a strong link to the stakeholder need. Secondly, it enables us to implement fast feedback loops for the adaptation of test specifications, requirements specifications, or the requirements model.

\circled{E} 
In order to foster the information exchange between requirements and test engineers and to increase the understanding of the system behavior, we added the sub-process~E. In E we use the \emph{SMLK animator} to simulate requirements based on the concepts of \emph{LSC Play-Out} \cite{Harel2003} and \emph{scenariotools}~\cite{Greenyer2017}, as introduced in Sect.\,\ref{sect:smlk}. For the requirements simulation, we use two input artifacts:~1) the test scenarios, that we create by running through the sub-process B, C, and D, and 2) the specification scenarios that we create by applying the TDSS process in C. 

By step-wise executing the requirements, driven by the test scenarios, the interaction between single components is visualized. These visualizations can be used to refine requirements and test scenarios, or to discuss and align the stakeholder expectations. Hence, we use the sub-process D to directly execute the features and get fast feedback to iteratively refine test and specification scenarios. As part of the sub-process E we execute and visualize these scenarios to cross-check if the modeled behavior meets the expectations. If the participating roles agree on test and system behavior, the generated sequence diagrams can be used to support the targeted artifacts (requirements and integration test specification) of our approach. 
 
\section{Case Study}
\label{sect:casestudy}
To gain insight into the practical implications of our approach for future ECU development projects, we conducted a case study at a Tier1 supplier company. We publish our used tools on GitHub\footnote{\url{https://github.com/qualicen/specmate} for the test case generation in B} and BitBucket
\footnote{\url{https://bitbucket.org/jgreenyer/smlk/} for the scenario-based formalization of functional requirements in C } 
\footnote{\url{https://bitbucket.org/crstnwchr/besos/} for independent modeling of component interactions and component behavior in C. This project also includes the Cucumber tool to support the BDD approach in D}
\footnote{\url{https://bitbucket.org/jgreenyer/smlk-animator/} to simulate the component interaction and generate sequence diagrams in E} to allow others to evaluate and evolve the single process steps. 
Following the Goal-Question-Metric~\cite{GQM} technique, we define the goal of our case study as follows:

\begin{enumerate}
    \item \textit{Object}. A requirements specification that links stakeholder requirements with component requirements, and a test specification to validate the component interactions.
    \item \textit{Purpose}. Evaluate to what extent the artifacts of our integrated requirements analysis and test specification approach can improve the existing specifications at the case company.   
    \item \textit{Focus}. Specification of requirements and design of test scenarios.  
    \item \textit{Viewpoint}. Software architects that derive requirements from stakeholder requirements, and test designers that create integration test specifications. 
    \item \textit{Context}. Software development within an ASPICE compliant ECU development project. 
\end{enumerate}

The expected outcome of our study is an evaluation of whether our approach can extend or improve existing requirements and test specifications by providing \emph{Artifact 1} and \emph{Artifact 2} (see Fig.~\ref{fig:method}).  In the following, we apply our top-down approach as well as the bottom-up approach and outline how our method supports practitioners in the specification and testing of integrated system behavior.

\subsection{Background of the Case Study}
\label{sect:casestudyApproach}
We performed the individual process steps in collaboration with our industry partner\footnote{\url{https://www.kostal.com/en-gb}} based on a complex real-world application. Specifically, we focused on the \emph{plug interlock} function, which is responsible for locking the connected charging plug to prevent it from being disconnected during an active charging process. To gather feedback about the applicability of our approach, we involve 3 experts in the execution of the top-down approach and 6 experts in the execution of the bottom-up approach. 

According to the automotive safety integrity level (ASIL) classification scheme (see ISO26262 \cite{ISO26262_2018}), the \emph{plug interlock} function is classified as ASIL A. Hence, a comprehensive test and requirements specification is required. At our case company, this function was developed in two locally distributed teams, where one team was responsible for developing the low-level hardware logic for the locking motor actuation, and one team developed the evaluation logic for the single signals coming from the charging infrastructure. Based on the evaluated signal information, a third component with a high-level application logic was provided by an external company. This component was integrated based on an interface specification. The detailed component behavior was not known. Hence, requirements and component tests for the application logic were not specified at the Tier1 supplier company, but this component was integrated with the other components in one ECU software system. Therefore, it was necessary to consider the signals of this external application logic for integration testing of the plug interlock feature as well. 

\subsection{Application: top-down integration strategy}

\textbf{Entry Point D}
We started with the specification of acceptance criteria, by defining features (D1) including usage scenarios (D2), as exemplified in Listing~\ref{list:feature}.   
Based on the feature specifications, we generated test skeletons (D3) as shown in Listing \ref{list:testSkeleton}. 
Subsequently, we used these test skeletons to drive the modeling of inter-component scenarios. 
\begin{lstlisting}[caption=Feature specification with two usage scenarios based on the\\ Gherkin syntax \cite{SmartBearSoftware},
	label=list:feature,
	style=GherkinStyle
	]
Feature: Plug Interlock
  Scenario: the user connects plug to socket A
    Given the user connects plug A to socket A
    When socket A sends the pilot signals to the OBC
    Then the OBC shall detect and interlock the plug A
\end{lstlisting}
The \textbf{central artifacts} of the sub-process D are the \emph{feature specification} (Listing \ref{list:feature}) and the generated \emph{test skeletons} (Listing \ref{list:testSkeleton}).    

\begin{figure}[h]
    \centering
    \includegraphics[width=1.0\linewidth]{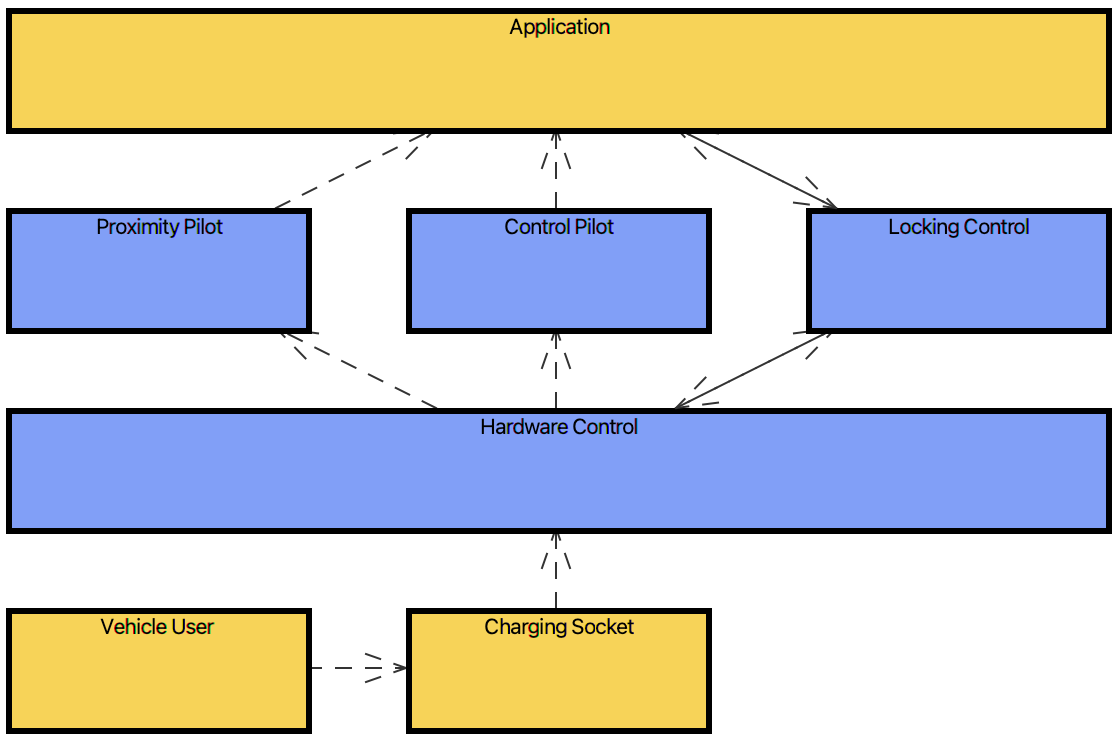}
    \caption{Visualization of the component interaction in the SMLK animator}
    \label{fig:simulation}
\end{figure}

\textbf{Entry Point C}
To specify the inter-component scenarios, we used the previously generated test skeletons and started with the TDSS process in C to iteratively specify the necessary components and their interaction. Based on existing software architecture specifications of previous development projects and the available high-level OBC system specification, we identified the components shown in Fig.\,\ref{fig:simulation} to realize the plug interlock feature: 
The \emph{vehicle user} and the \emph{charging socket} are external systems, where the charging socket includes the motor that shall be actuated by the OBC. The \emph{hardware logic}, \emph{proximity pilot}, \emph{control pilot}, and \emph{locking control} represent the requirements in development. The \emph{application} represents the external to be integrated software.  
\begin{figure*}[h]
    \centering
    \includegraphics[width=1.0\linewidth]{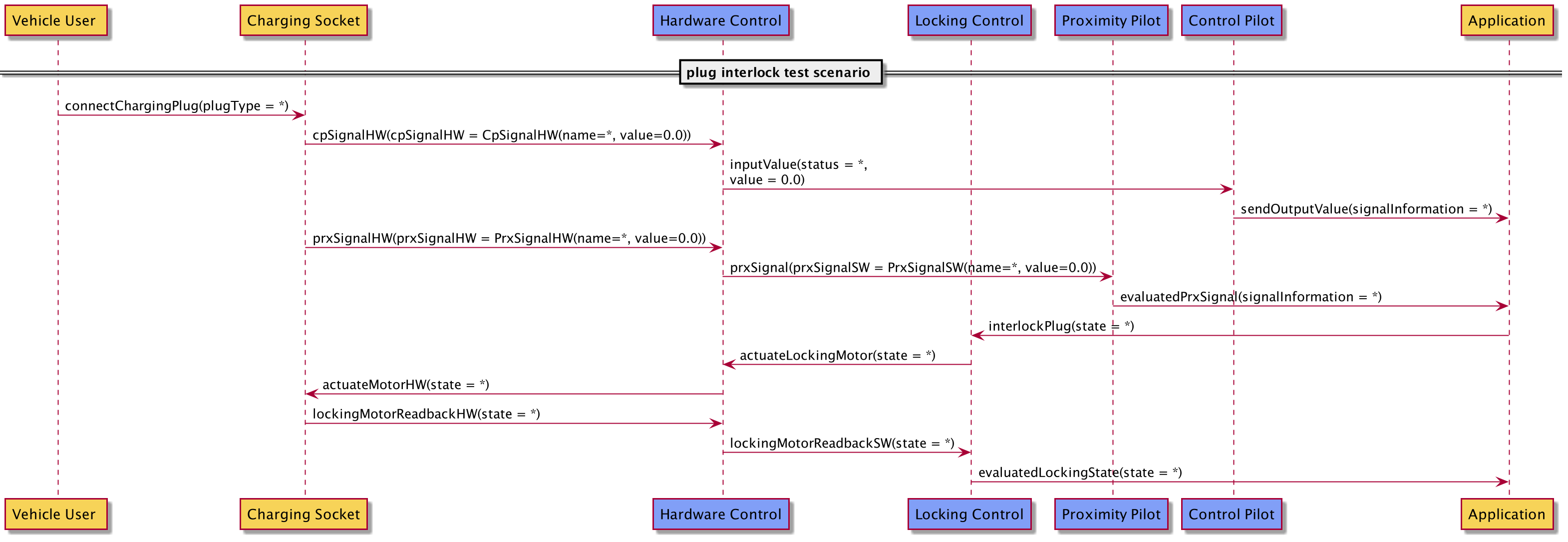}
    \caption{Generated sequence diagram as result of the requirements simulation in sub-process E.}
    \label{fig:sequence}
\end{figure*}

After these components were set, we started with iterating in C. We first focused on the sequence in which information should be exchanged without providing details for specific signal values.
\begin{lstlisting}[caption=Generated test skeletons and manually added SMLK events,
	label=list:testSkeleton,
	style=GherkinStyle
	]
Given("the user connects plug A to socket A"){
    trigger(vehicleUser sends chargingSocket.connectChargingPlug("plug A"))} //manually added 
When("the charging socketA sends the pilot signals to the OBC"){
    trigger(chargingSocket sends hardwareControl.cpSignalHW(CpSignalHW("*", 0.0))) //manually added
    trigger(chargingSocket sends hardwareControl.prxSignalHW(PrxSignalHW("*", 0.0)))} //manually added  
Then("the OBC shall detect and interlock the plug A"){
    eventually(hardwareControl sends chargingSocket.actuateMotor())} //manually added
\end{lstlisting}
Although we abstract from component details, we were already able to execute the component interaction by applying the modeling concepts introduced in Sect. \ref{sect:smlk}.
Accordingly, we triggered the execution of inter-component scenarios by external events that we manually added to the generated test skeletons. 
E.g., in Listing~\ref{list:testSkeleton} we specified the expected system behavior in the case of a user who connects a charging plug (\textcolor{darkviolet}{Given}). When the pilot signals were sent (\textcolor{darkviolet}{When}), we expect the hardware controller to actuate the locking motor (\textcolor{darkviolet}{Then}). In this example, the message \lstinlineKotlin{actuateMotor} is only sent back from the \lstinlineKotlin{HardwareControl} to the \lstinlineKotlin{ChargingSocket} if this is specified as an inter-component scenario as done in Listing~\ref{list:interComponentScenario2}. 
\begin{lstlisting}[caption=Inter-component scenario to drive the locking motor,
	label=list:interComponentScenario2,
	style=KotlinStyle
	]
scenario(application sends lockingControl receives LockingControl::interlockPlug){
    request(lockingControl sends hardwareControl.actuateLockingMotor("*"))
    request(hardwareControl sends chargingSocket.actuateMotorHW("*"))
    ...}
\end{lstlisting}

In this way, we were able to drive the scenario modeling through tests that we derived from the feature specification.       
Thereby, we consider the complete software system of the ECU as a black-box to describe the end-to-end event chain (see Fig.~\,\ref{fig:sequence}). Hence, we also considered interactions with the externally developed software without having access to detailed specifications. However, by using inter-component scenarios, we were able to model the interaction based on the given interface specification\footnote{Due to confidentiality restrictions, details of the specification cannot be shown. Nevertheless, we provide an overview of the signal names, the data types, and their intended use in this section.}, and driven by the high-level feature specification (Listing~\ref{list:feature}). 

After we successfully executed the feature, we started with the specification of the test scenario as our central artifact (see Listing~\ref{list:testScenario}). For this purpose, we used the information from the enriched test skeletons (artifact 2d) and build on the events that are sent and received by the external components (vehicle user and charging socket). In lines 3 - 5 we model that the vehicle user connects a charging plug, and we request the pilot signals. In line 9 we wait for the message \lstinlineKotlin{actuateMotorHW}, indicating that the \lstinlineKotlin{HardwareControl} wants to lock the plug. Additionally, since the goal of the test scenario is to validate the requirements in development, we also considered the interaction with the application in this test scenario. Therefore, we used the previously modeled end-to-end chain (see Fig.~\ref{fig:sequence}), to identify which messages are received and sent by the application. This lead to line 6 and 7, where we wait for the pilot signals to be set, and line 8, where we model that the application responds with a locking request. 
\begin{lstlisting}[caption=Test scenario,
	label=list:testScenario,
	style=KotlinStyle
	]
scenario{
label("plug interlock test scenario")
request(vehicleUser sends chargingSocket.connectChargingPlug())
request(chargingSocket sends hardwareControl.cpSignalHW())
request(chargingSocket sends hardwareControl.prxSignalHW())
waitFor(proximityPilot sends application receives Application::evaluatedPrxSignal)
waitFor(controlPilot sends application receives Application::evaluatedCpSignal)
request(application sends lockingControl.interlockPlug())
waitFor(hardwareControl sends chargingSocket.actuateMotorHW())}
\end{lstlisting}
The \textbf{central artifacts} of the sub-process C are the \emph{test scenarios} (Listing \ref{list:testScenario}) and a \emph{scenario specification} including inter-component scenarios (Listing \ref{list:interComponentScenario2}).

\textbf{Entry Point E} After we created the test scenario and the inter-component scenarios in C, we used these artifacts as input for the requirements simulating in E.
Triggered by the test scenarios, we were able to step-wise execute each interaction of the end-to-end chain to validate the modeled behavior and to subsequently generate the \emph{sequence diagram} as shown in Fig.~\,\ref{fig:sequence}.
as the \textbf{central artifact} of the top-down integration strategy. 

\subsection{Application: bottom-up integration strategy}
\label{sect:casestudySpecmate}
\textbf{Entry Point B}
After we specified a first test scenario for the high-level component interaction, we continued with the bottom-up approach in B. 
Therefore, we used the specification of the control pilot component. This component is responsible for pre-processing the control pilot signal, in order to provide an evaluated information to the application logic (see the component structure in Fig.\,\ref{fig:simulation} which is used to specify the plug interlock feature).  

\begin{figure}[h]
    \centering
    \includegraphics[width=0.7\linewidth]{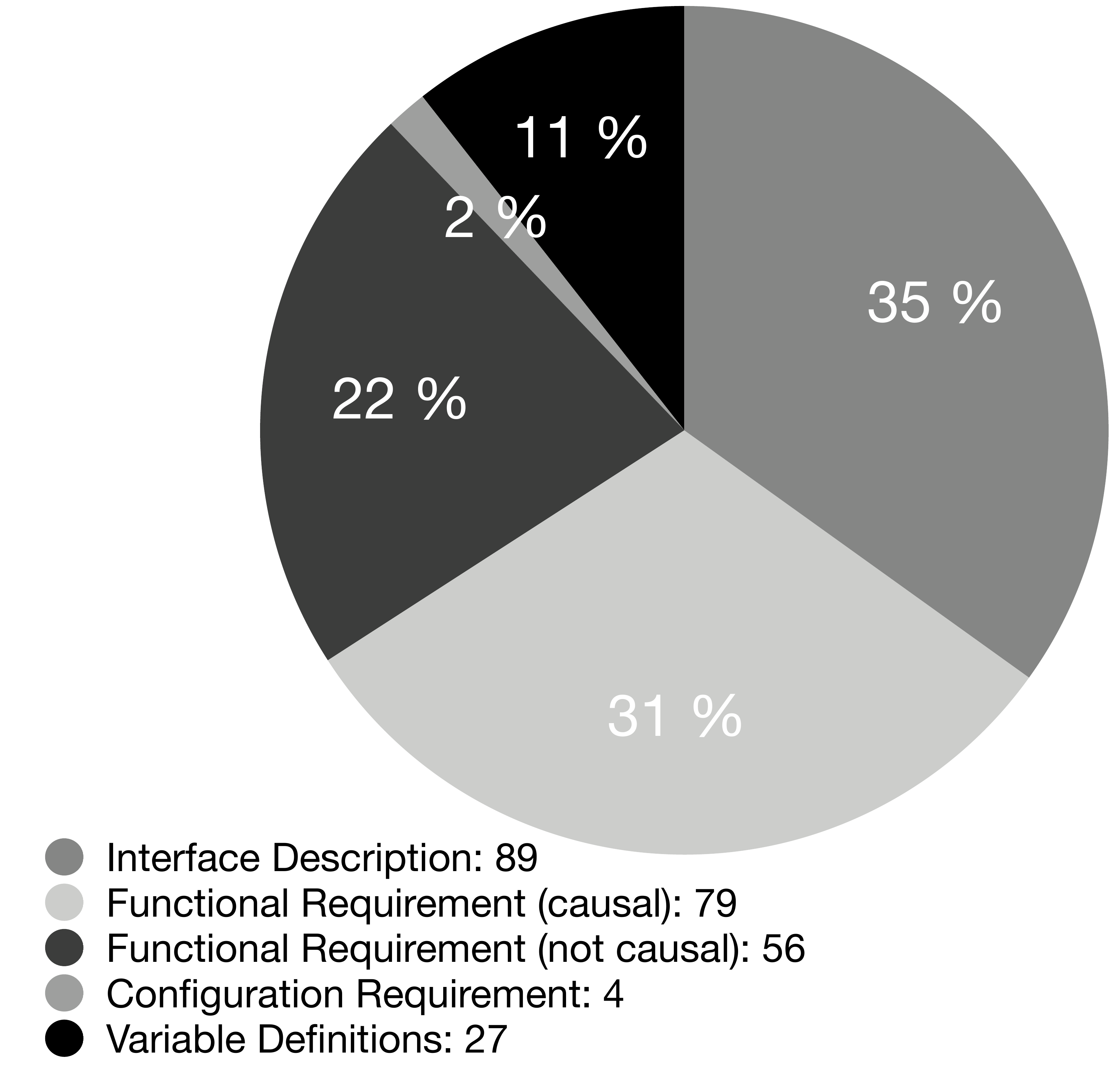}
    \caption{Categories of requirements of the used specification.}
    \label{fig:reqTypes}
\end{figure}

As shown in Fig.~\ref{fig:reqTypes}, the investigated specification includes 255 requirements. Out of these 255 requirements, 135 requirements are functional and specify the expected system behavior. In addition to these functional requirements, 89 requirements define interface names and provide basic information about the intended interface usage. 4 requirements provide information on how to configure the component. The remaining 27 requirements include information on how variables should be defined. These non-functional requirements are not in our scope. In our case study, we focus on functional requirements that follow a causal pattern, making them processable for \emph{Specmate}. Our analysis shows that 79 functional requirements are indeed causal, while 56 requirements describe the functional behavior in a static way. E.g.: "\textit{The signal \texttt{signalName} shall be set to \texttt{InitValue}}". With adding more information to these 56 requirements (e.g., by providing more information about when the signal should be set), they could also be considered by our approach. 
 
\begin{figure}
    \centering
\resizebox{\columnwidth}{!}{%
\begin{tikzpicture}
\def\radius{2cm}
\def\mycolorbox#1{\textcolor{#1}{\rule{2ex}{2ex}}}
\colorlet{colori}{black!70}
\colorlet{colorii}{black!50}

\coordinate (ceni);
\coordinate[xshift=\radius] (cenii);

\draw[fill=colori,fill opacity=0.5] (ceni) circle (\radius);
\draw[fill=colorii,fill opacity=0.5] (cenii) circle (\radius);

\draw  ([xshift=-20pt,yshift=20pt]current bounding box.north west) 
  rectangle ([xshift=20pt,yshift=-20pt]current bounding box.south east);


\node at ([xshift=\radius]current bounding box.east) 
{
\begin{tabular}{@{}lc@{}}
& \\
\mycolorbox{colori!50} & Manual Set \\
\mycolorbox{colorii!50} & Automatic Set \\
\mycolorbox{black!70!black!50} & Overlapping  \\
\end{tabular}
};

\node[xshift=-.5\radius] at (ceni) {$59$};
\node[xshift=.5\radius] at (cenii) {$22$};
\node[xshift=.9\radius] at (ceni) {$145$};
\end{tikzpicture}%
}\caption{Comparison of manually and automatically created test cases by \emph{Specmate}.}
    \label{fig:testSuiteCompare}
\end{figure}
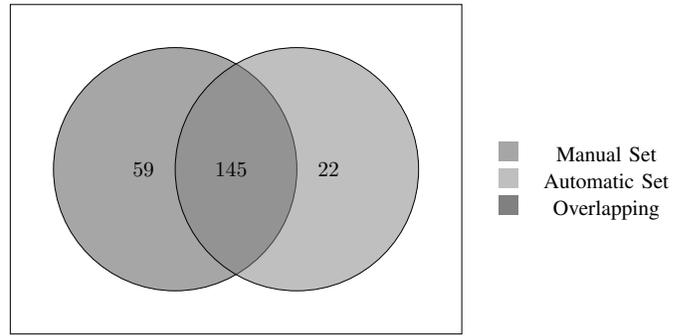

Based on the 79 causal requirements, we investigated the potential to automatically generate test cases as input for the TDSS process. To this end, we evaluate each step in the \emph{Specmate} pipeline (see B in Fig.\ref{fig:method}) and investigate the following research questions (RQ):

\textbf{RQ1:} Can \emph{Specmate} generate the same test cases as the conventional manual approach?
The first research question investigates whether \emph{Specmate} can achieve the status quo. To answer this question, we apply \emph{Specmate} to the 79 causal requirements and compare the created test cases with the manual set. Specifically, we examine three subsets: the first set contains the intersection of both test case sets, while the second set consists of test cases created exclusively by the manual approach. The last subset consists exclusively of automatically created test cases (see Fig.~\,\ref{fig:testSuiteCompare}).

\textbf{RQ2:} What are the most common causes for deviating test cases?
With the second RQ, we want to determine why certain test cases could only be created manually or only automatically. This provides us an insight into the further optimization potential of \emph{Specmate}.

\paragraph{Answer for RQ 1}
Interestingly, the manual test specification does not contain corresponding test cases for all causal requirements. For five of the 79 causal requirements, no test cases were derived manually. In total, the manual test specification contains 204 test cases, which corresponds to about 2.58 test cases per requirement. By using \emph{Specmate}, we were able to create test cases for 68 of the 79 causal requirements. In total, the automated generated test set contains 167 test cases. 
Comparing the manual set with the automated set, it is evident that the majority of the manually created test cases could also be created automatically by \emph{Specmate}. In total, 145 of the manual test cases can be found in the automated set, corresponding to an overlapping rate of about 71\%. However, 59 of the manually created test cases could not be created automatically. Nevertheless, \emph{Specmate} was able to create 22 test cases that had not previously been considered in the manual process. According to the test designers of our industry partner, these test cases are technically correct and should be included in the test suite. Hence, \emph{Specmate} allowed us to detect errors in the manual testing process.

\paragraph{Answer for RQ 2}
\label{sect:specmateRq}
After the quantitative comparison of both test sets, we analyzed why some test cases could only be created manually or automatically. We found the following reasons:

\begin{enumerate}
    \item  Our study shows the lack of robustness of \emph{Specmate} to grammatical errors in a sentence. In eight of the causal sentences, commas were forgotten or placed in the wrong position. Hence, an incorrect dependency parse tree was generated from which \emph{Specmate} could not create a suitable CEG. We also encountered problems with handling complex variable names, since they were not included in the vocabulary on which the MaltParser was trained. As a result of these two problems, 32 of the 59 exclusively manually created test cases could not be covered by \emph{Specmate}.
    \item The other 27 missing test cases are caused by the fact that nine of the causal requirements do not contain all the information needed to create all necessary test cases. According to the test designers of our industry partner, in such cases, domain knowledge is required to create all necessary test cases despite the lack of adequate descriptions. This stresses the strong reliance of our automated approach on the quality of the requirements documentation.
    \item 13 of the 22 test cases, which were only created automatically, refer to the five causal requirements for which no test cases were created manually.  The remaining nine test cases stem from ignoring negative testing. We hypothesize that there is a tendency to first create test cases that check the desired system behavior on the basis of valid inputs (positive testing). This causes negative test cases to be missing. 
    \end{enumerate}
    
As shown in the process overview (Fig.~\,\ref{fig:method}), when following the bottom-up approach, we used feedback loops to add information to requirements (artifact 1a) and correct grammatical errors within the specification, based on the findings in \ref{sect:specmateRq}. 

\textbf{Entry Point C}
Subsequently, based on the set of generated test cases (artifact 2a), 
we again iterated in C, but now we modeled \emph{component scenarios} driven by the generated tests. 
To guide the modeling of component scenarios, we used the previously defined end-to-end event chain (artifact 2e) to decide which component requirements are required when implementing the sequence. 
As an example, for the control pilot component, the requirements describing the processing of the signal values were relevant within the event chain. Therefore, we selected all generated tests that were generated from these requirements as input for the component-level TDSS. 
\begin{figure}[h]
    \centering
    \includegraphics[width=1.0\linewidth]{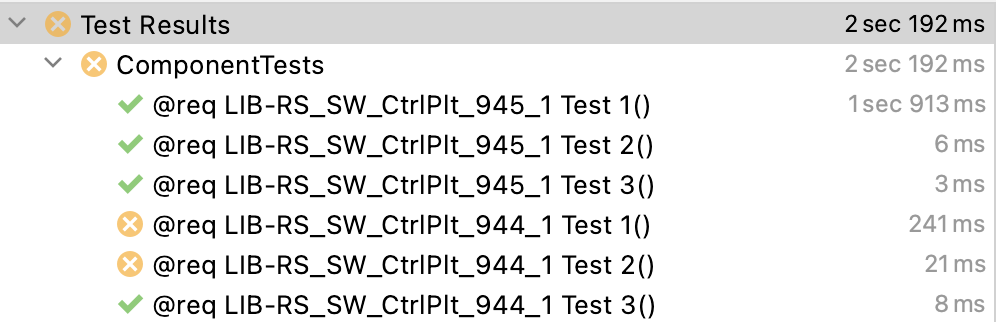}
    \caption{Component tests that were used to drive the component scenario modeling}
    \label{fig:testResultsComponent}
\end{figure}
Fig.~\,\ref{fig:testResultsComponent} exemplary shows test results of one TDSS iteration, where we see two requirements (from 15 identified requirements) with three tests for each requirement. 

In this way, the results of the requirements simulation (artifact 2e), as part of the top-down integration strategy, support the identification of required component level requirements. And consequently, we were able to focus on the right amount of requirements (12 for the control pilot component) to be modeled, driven by 35 generated tests. 
 
\subsection{Application: Interlinking both Strategies}
As a next step, we were interested in how the test scenario (artifact 2b) could be refined when integrating the inter-component scenarios with the previously created component scenarios. Therefore, we tested the integrated behavior by jointly executing the single scenario programs, triggered by the feature shown in Listing \ref{list:feature}.
\begin{figure}[h]
    \centering
    \includegraphics[width=1.0\linewidth]{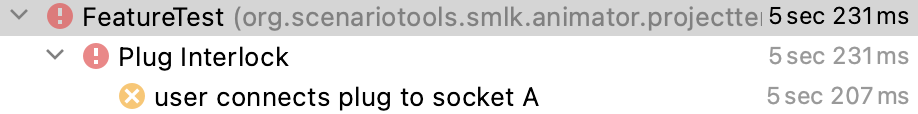}
    \caption{Combined feature with considering component details}
    \label{fig:testResultsCombined}
\end{figure}
This led a failed test result as shown in Fig.~\,\ref{fig:testResultsCombined}. By again stepping through the simulation in E we found that the \lstinlineKotlin{ControlPilotObject} now expects the signal status \lstinlineKotlin{RTE_E_OK} and the signal value \lstinlineKotlin{3.0} as shown in the component scenario in Listing \ref{list:intraComponentScenario}. Consequently, we used this information to extend our test scenario in line 4 with this exact signal value to repair our feature test. 
This allowed us to insert details from the component specification into the test scenario we derived from the high-level component interaction.


\section{Lessons Learned}

This section provides a summary of the lessons learned that emerged from our case study. Specifically, we describe the practical impact of our approach and the resulting take-aways for practitioners.

\textbf{Lesson 1} \textit{Our generated sequence diagrams (artifact 1b) are well suited to create requirements specifications}: 

Our industry partner reported that the implemented development processes require the specification of component interactions, e.g., in the form of UML sequence diagrams. However, according to the experts, the manual creation and maintenance of these sequence diagrams were not feasible for large applications, since stakeholder requirements and component behavior frequently change. Consequently, the manually created sequence diagrams were often outdated and not compliant with the implementation. This poses a major problem, since these sequence diagrams are part of the test basis for the integration test design, and need to be up-to-date and correctly specified. We address this problem and improve the state of practice because we generate the sequence diagrams automatically in E after simulating the component interactions. Instead of manually specifying sequence diagrams as a test basis, we drive the requirements modeling by the tests and generate the sequence diagrams as a result. 

\textbf{Lesson 2} \textit{Our approach promotes detailed modeling}:

To increase the feasibility of requirements modeling, it is necessary to decide to what extent details should be considered. We found that our combined top-down and bottom-up integration strategy can guide practitioners to identify component requirements that are linked to stakeholder requirements of a specific feature in development, and therefore should be considered for modeling.   

\textbf{Lesson 3} \textit{Automated analysis of component interactions and component behavior helps to identify the potential of component reuse}:  

Especially for automotive development projects, where functional safety of systems must be confirmed with high verification and validation effort, component reuse is important to reduce costs. We found that our approach can serve as a tool to give a first indication where requirements must be updated when stakeholder requirements change. E.g., to answer the questions if the existing specification is still valid if new tolerance values must be considered.      

\textbf{Lesson 4} \textit{Automated test case generation can improve existing component test processes:}

Based on the results in Sect. \ref{sect:casestudySpecmate}, the automated generation of test cases was able to improve the component test design process at our case company. Even if the generated test cases are not used as input for our integrated analysis and specification approach, the sub-process B of our approach is suited to generate component tests that verify the implemented component. 
Specifically, since we compared the generated test cases with an existing manually created component test specification\footnote{created with the testing tool TESSY: \url{https://www.razorcat.com/en/product-tessy.html}}, and we found that the majority of test cases can be generated, we can use these generated test cases as input for testing the implemented program code. 

\textbf{Lesson 5} \textit{Iterations and feedback loops can increase the feasibility of requirements modeling and validation in practice}: 

The consideration of the different entry points (A-E) is motivated by the state of practice where we do not have complete models, data sets, or specifications. Therefore, depending on the available data, we can start with specifying new requirements (A), parse existing requirements (B), or start with the specification of high-level features (D). This increases the feasibility of requirements modeling in practice, since our approach does not depend on formally correct input data and complete system specifications. Instead, we start with informal stakeholder information or component requirements and iteratively create the scenario-based requirements models. Thereby, driven by the tests, we get fast feedback and can iteratively extend the scenario specifications. 

\textbf{Lesson 6} \textit{The created test scenario (artifact 2a) and the generated sequence diagrams (artifact 2e) are well suited to extend existing integration test specifications:}

As shown in Sect. \ref{sect:casestudyApproach}, we modeled the component interaction driven by acceptance tests. 
With using inter-component scenarios, we were able to model complete end-to-end event chains, by also considering the interaction with external software components.
According to the experts at our case company, this information about end-to-end behavior is valuable for designing integration test scenarios.   
Especially for the development and integration of software components in distributed development environments, as this was the case for the \emph{plug interlock} function used in our case study.
In particular, the generated sequence diagrams, which visualize the interaction between the test scenario and the modeled requirements, are suitable for extending existing integration test specifications.  






\section{Threats to Validity}
As in every empirical study, our case study is also subject to potential validity threats. The internal validity of our study might be subject to a selection bias, since we selected our case study partner based on convenience sampling. In addition, the comparison between the manually and automatically created test cases by \emph{Specmate} might be subject to researcher bias. To mitigate this risk, the first and second author individually mapped the test cases. Subsequently, the mapping was cross-checked by the experts and discussed within the research group. 
Our greatest concern for external validity arises from the fact that we evaluated our approach only in the context of a single automotive development project. Moreover, we apply our approach only to small set of existing requirement and test specifications. This small study sample allowed us to execute and demonstrate our approach, however, we can not provide insights on the scalability of our approach. In addition, our evaluation results presented above might not be generalizable to all other situations and contexts. However, we argue that other domains (e.g., insurance, aerospace) face also the challenge of specifying and testing an integrated system behavior consisting of multiple components. Hence, we assume that our idea of combining a top-down approach with a bottom-up approach is well suited and leads to an increasing support for practitioners across domains.

\section{Closing and Next Steps}
Currently, practitioners follow a top-down approach in automotive development projects. However, recent studies~\cite{Kirpes2019,Juhnke2020,Kasauli2021} revealed that this top-down approach is not suitable for the implementation and testing of modern automotive systems. Practitioners increasingly tend not to specify requirements and tests for systems with complex component interactions (e.g., e-mobility systems). We address this research gap and present an integrated and iterative scenario-based technique for the specification of requirements and test scenarios. Our approach combines both a top-down and a bottom-up integration strategy. As a top-down approach, we build on our scenarios in the loop (SCIL) approach~\cite{Wiecher2020} where we use early requirements prototyping to align stakeholder needs with technical requirements. In this paper, we extend SCIL with a bottom-up integration approach and specifically focus on the specification of tests and requirements. Therefore, we combined the NLP technique~\cite{Fischbach2020} with scenarios-based modeling techniques to specify the interaction between components, while also including component-level requirements.

We have shown that our approach has the potential to support software architects and test designers in the integrative specification of cross-component requirements and corresponding integration tests. 
In future work, we will investigate more practical implications of our approach by integrating it into upcoming development projects in the e-mobility context. 


\bibliographystyle{ieeetr}
\bibliography{references}
\end{document}